\documentclass[preprint,aps,pra,superscriptaddress]{revtex4-1}
 \usepackage{amsmath}
\usepackage{amssymb}
\usepackage{graphicx}
 \usepackage{color}
  \usepackage{hyperref}
\usepackage{subfigure}
\usepackage[normalem]{ulem}

{\vskip 2pt \begin{compactitem}[#1]\setlength{\itemsep}{2pt}}
{\end{compactitem}\vskip 2pt}

\newcommand{\be}{\begin{equation}}
\newcommand{\ee}{\end{equation}} 
\newcommand{\lb}{\label}
\newcommand{\OL}{\overline}

\newcommand{\ba}{{\bf a}}

\newcommand{\bk}{{\bf k}}

\newcommand{\br}{{\bf r}}
\newcommand{\bu}{{\bf u}}

\newcommand{\bx}{{\bf x}}

\newcommand{\wt}{\widetilde}

\newcommand{\grad}{{\mbox{\boldmath $\nabla$}}}
\newcommand{\bdot}{{\mbox{\boldmath $\cdot$}}}

\newcommand{\btimes}{{\mbox{\boldmath $\times$}}}

\begin{document}

\title{\textbf {The Inviscid Criterion for Decomposing Scales}}
\author{Dongxiao Zhao}
\email{dzhao5@ur.rochester.edu}
\affiliation{Department of Mechanical Engineering, University of Rochester}
\author{Hussein Aluie}
\affiliation{Department of Mechanical Engineering, University of Rochester}
\affiliation{Laboratory for Laser Energetics, University of Rochester}

\begin{abstract}
The proper scale decomposition in flows with significant density variations is not as straightforward as in incompressible flows, with many possible ways to define a `length-scale.' A choice can be made according to the so-called \emph{inviscid criterion} \cite{Aluie13}. It is a kinematic requirement that a scale decomposition yield negligible viscous effects at large enough `length-scales.'
It has been proved \cite{Aluie13} recently that a Favre decomposition satisfies the inviscid criterion, which is necessary to unravel inertial-range dynamics and the cascade. Here, we present numerical demonstrations of those results. We also show that two other commonly used decompositions can violate the inviscid criterion and, therefore, are not suitable to study inertial-range dynamics in variable-density  and compressible turbulence. Our results have practical modeling implication in showing that viscous terms in Large Eddy Simulations do not need to be modeled and can be neglected.
\end{abstract}

\maketitle

\section{Introduction}
The notion of a `length-scale' in a fluid flow does not exist as an independent entity but is associated with the specific flow variable being analyzed 
\footnote{The length-scale associated with velocity, ${\bf u}({\bf x})$, and that with vorticity, ${\bf \nabla \times u}({\bf x})$ for example, will be the same if the scale decomposition (e.g. Fourier transform) commutes with spatial derivatives.}. 
While this might seem obvious, we often discuss the `inertial range' or the 'viscous range' of length-scales in turbulence as if they exist independently of a flow variable, which in incompressible turbulence is the velocity field as Kolmogorov showed \cite{Kolmogorov41a}. The overarching theme of this paper pertains to the following question: \emph{Can an inertial-range exist for one quantity but not another within the same flow?} The answer is \emph{yes}.  Herring et al. \cite{herring1994ertel} studied the dynamics of a passive scalar, $\theta(\bx)$, advected by an incompressible turbulent velocity, $\bu(\bx)$, and showed that potential vorticity, $(\grad\btimes\bu)\bdot\grad\theta$, which is an ideal Lagrangian invariant, does not have an inertial range \footnote{The potential vorticity regime studied by \cite{herring1994ertel}, in the absence of rotation or stratification, is not geophysically relevant. As the authors of \cite{herring1994ertel} remark, in the presence of either rotation or stratification, potential vorticity is a predominantly linear quantity that has an inertial range and can therefore cascade.}. This is despite the existence of an inertial range for each of $\bu$ and $\theta$. They showed that this is due to significant viscous contributions to the evolution of $(\grad\btimes\bu)\bdot\grad\theta$ at \emph{all} length-scales, thereby precluding the existence of an inertial range.

In turbulent flows where significant density variations exist, we will show here that a similar situation can occur. 
In such flows, ascribing a length-scale to momentum or kinetic energy is not as straightforward as in incompressible
flows. Such quantities are one order higher in nonlinearity compared to their incompressible counterparts due to the density
field. This has led to different scale decompositions being used in the literature. `Length-scale' within these different decompositions correspond to different flow variables, each of which can yield quantities with units of momentum and energy. Ref. \cite{Aluie13} introduced an \emph{inviscid criterion} for choosing a proper decomposition to analyze inertial range dynamics and the cascade in such flows. The \emph{inviscid criterion} stipulates that a scale decomposition should guarantee a negligible contribution from viscous terms in the evolution equation of the large length-scales. Here, a length-scale is `large' relative to the viscous scales.  Based on this criterion, Ref. \cite{Aluie13} proved mathematically that a Hesselberg \cite{Hesselberg26} or Favre \cite{favre1958further,favre1969statistical} decomposition (hereafter, Favre decomposition) of momentum and kinetic energy satisfies the inviscid criterion, and then went on to show how an inertial range and a cascade \cite{Aluie11c,Aluieetal12,Aluie13} can exist in variable density high Reynolds number flows. However, Ref. \cite{Aluie13} did not prove the uniqueness of the Favre decomposition in satisfying the inviscid criterion, giving only physical arguments on why other decompositions used in the literature are expected to have significant viscous contamination at large length-scales and, therefore, are not suitable to study inertial-range dynamics.

Areas of application span many engineered and natural flow systems that have considerable density differences.
Large density ratios are often encountered in astrophysical systems, such as 
in molecular clouds in the interstellar medium which  have density ratios ranging from $10^6$ to $10^{20}$ (e.g. \cite{Kritsuketal07,Federrathetal10,Panetal16}). Much higher ratios can be expected in flow systems with gravitational effects, which can lead to the accretion of matter and the formation of ultra-dense protostars and protoplanets (e.g. \cite{HennebelleFalgarone12}). In high energy density physics (HEDP) applications performed at national laboratory facilities, such as in inertial confinement fusion (ICF) experiments, density ratios upward of $10^4-10^5$ are frequently encountered (e.g. \cite{Craxtonetal15,Yanetal16,LePapeetal16}).
In laboratory flow experiments, density ratios of up to 600 have been achieved using different fluids \cite{Read84,DimonteSchneider00}. Probably the most ubiquitous terrestrial two-fluid mixing is between air and water which have a density ratio of 1000. A systematic and rigorous scale-analysis framework is essential to understanding and modeling the mutliscale physics of such flows.

In this paper, we shall (i) present numerical demonstration that the Favre decomposition indeed satisfies the inviscid criterion, and (ii) that two other decompositions used in the literature do not satisfy the criterion. The results herein apply to flows with variable density due to compressibility effects and also to flows of incompressible fluids of different densities. In flows of the second type, which have been called ``variable density flows'' in the literature (e.g. \cite{Rayleigh1883,Sandoval95,Sandovaletal97,LivescuRistorcelli07}), density is not a thermodynamic variable and acoustic waves are absent. To simplify the presentation, we use the term `variable density' in this paper in reference to both types of flows.

\section{Decomposing Scales}
`Coarse-graining' or `filtering' provides a natural and versatile framework to understand scale interactions (e.g. \cite{Leonard74,MeneveauKatz00,Eyink05}). For any field $\ba(\bx)$, a coarse-grained or (low-pass) filtered field, which contains modes
at scales $>\ell$, is defined in $n$-dimensions as
\be
\OL \ba_\ell(\bx) = \int d^n\br~ G_\ell(\br)\, \ba(\bx+\br),
\lb{eq:filtering}\ee
where $G(\br)$ is a normalized convolution kernel and $G_\ell(\br)= \ell^{-n} G(\br/\ell)$ is a dilated version of the kernel having its main support over a region of diameter $\ell$. The framework is very general and includes Fourier analysis (e.g. \cite{Krantz99,AluieEyink09}) and wavelet analysis (e.g. \cite{meneveau1991analysis,meneveau1991dual}) as special cases with the appropriate choice of kernel $G(\br)$. The scale decomposition in (\ref{eq:filtering}) is essentially a partitioning of scales in the system into large ($\gtrsim\ell$), captured by $\OL \ba_\ell$, and small ($\lesssim\ell$), captured by the residual $\ba'_\ell=\ba-\OL \ba_\ell$. More extensive discussions of the framework and its utility can be found in many references (e.g. \cite{Piomellietal91,Germano92,Meneveau94,Eyink95,Eyink95prl,Chenetal03,AluieEyink10,FangOuellette16,Aluie17}). In what follows, we shall drop subscript $\ell$ when there is no risk for ambiguity.

In incompressible turbulence, our understanding of the scale dynamics of kinetic energy, such as its cascade, 
centers on analyzing $|\OL{\bu}_\ell|^2/2$. In the language of Fourier analysis, this is equivalent
to analyzing the velocity spectrum $|\widehat{\bu}(\bk)|^2/2$, where $\widehat{\bu}(\bk)$ is the Fourier transform of the velocity field $\bu(\bx)$ (see, for example, section 2.4 in \cite{Frisch95}).

In variable density turbulence, scale decomposition is not as straightforward. One possible
decomposition is to define large-scale kinetic energy as $\OL\rho_\ell |\OL\bu_\ell|^2/2$,
which has been used in several studies (e.g. \cite{chassaing1985alternative,BodonyLele05,Burton11,KarimiGirimaji17}). Another 
possibility is to define large-scale kinetic energy as $|\OL{(\sqrt{\rho}\bu)}_\ell|^2/2$,
which has also been used extensively in compressible turbulence studies
(e.g. \cite{kida1990energy,CookZhou02,Wangetal13,Greteetal17}). 
A third decomposition mostly popular in compressible 
large eddy simulation (LES) modeling uses $\OL\rho_\ell|\wt{\bu}_\ell|^2/2$ 
as the definition of large-scale kinetic energy, where
\be
\wt\bu_\ell(\bx) = \OL{\rho \bu}_\ell/\OL\rho_\ell~~.
\lb{eq:Favrefiltering}\ee
This decomposition was apparently first 
introduced by Hesselberg in 1926 \cite{Hesselberg26} to study stratified atmospheric flows, 
although it is often associated with Favre \cite{Favre58a,Favre58b,Favre58c,Favre58d} who first used it in 1958 to analyze compressible turbulence \footnote{These early studies used ensemble averaging or (Reynolds averaging) rather than filtering to decompose scales. There is a one-to-one correspondence in the definitions by replacing the ensemble average operation with the filtering operation wherever it appears.}. For a constant density, all these definitions reduce to the incompressible case.

It seems that there is no fundamental {\it a priori} reason to favor one 
definition over another. It has been argued that the Favre
decomposition is preferred from a ``fundamental physics'' standpoint
since it treats mass and momentum as the elemental variables. While this
is certainly a plausible justification, the argument does not identify precisely
what physics is missed when utilizing alternate decompositions. 

\section{The Inviscid Criterion\lb{sec:InviscidCriterion}}
In this paper, we will show that that the non-Favre decompositions can miss
the inertial-range physics if density variations are significant.  More precisely, 
we shall show that those alternate decompositions fail to satisfy the \emph{inviscid criterion}.

It is possible to derive the large-scale budgets governing each of those definitions, starting from the original
equations (\ref{continuity}),(\ref{momentum}) of continuity and momentum. Applying the filtering operation to the different combinations of density and velocity forming the three definitions of large-scale kinetic energy \cite{Aluie13}, one gets
\begin{eqnarray}
\partial_t \OL\rho_\ell|\wt{\bu}_\ell|^2/2 + \dots &=& \dots + \Sigma_\ell^F \lb{eq:FavreKEbudget}\\
\partial_t \OL\rho_\ell |\OL\bu_\ell|^2/2 + \dots &=& \dots + \Sigma_\ell^C \lb{eq:ChassaingKEbudget}\\
\partial_t |\OL{(\sqrt{\rho}\bu)}_\ell|^2/2 + \dots &=& \dots + \Sigma_\ell^K, \lb{eq:KritsukKEbudget} 
\end{eqnarray}
where the three viscous terms corresponding to the three definitions of large-scale kinetic energy are
\begin{eqnarray}
\Sigma_\ell^F &=& \wt{u}_i \, \partial_j \OL{\sigma}_{ij} = \underbrace{\partial_j\left[\wt{u}_i \, \OL{\sigma}_{ij}\right]}_{\Sigma_\ell^{F,\mbox{\scriptsize{diff}}}} -\underbrace{\left(\partial_j\wt{u}_i\right) \,  \OL{\sigma}_{ij}}_{\Sigma_\ell^{F,\mbox{\scriptsize{diss}}}}  ~~, \lb{eq:FavreVD}\\[9pt]
\Sigma_\ell^C &=& \OL{\rho}\,\,\OL{u}_i\,\OL{\left(\partial_j\sigma_{ij}\right)/\rho} \nonumber\\[6pt]
&=& \underbrace{\partial_j\left[\OL{\rho}\,\,\OL{u}_i\,\OL{\sigma_{ij}/\rho}\right]}_{\Sigma_\ell^{C,\mbox{\scriptsize{diff}}}} -\underbrace{\left[\left(\partial_j\OL{u}_i\right) \,\OL\rho\,\, \OL{\sigma_{ij}/\rho} +\OL{u}_i\left(\partial_j\OL{\rho}\right) \OL{\sigma_{ij}/\rho} +\OL{u}_i\,\, \OL{\rho}\,\, \OL{\sigma_{ij}\partial_j\left(1/\rho\right)}\right]}_{\Sigma_\ell^{C,\mbox{\scriptsize{diss}}}}, \lb{eq:ChassaingVD}\\[9pt]
\Sigma_\ell^K &=&  \OL{\sqrt\rho\, u_i}\,\OL{\left[\frac{1}{\sqrt\rho}\,\partial_j\sigma_{ij}\right]} \nonumber\\[6pt]
&=& \underbrace{\partial_j\left(\OL{\sqrt\rho u_i}\,\,\,\OL{\sigma_{ij}/\sqrt\rho}\right)}_{\Sigma_\ell^{K,\mbox{\scriptsize{diff}}}} 
-\underbrace{\left[\OL{\sqrt\rho\,\partial_j u_i} \,\,\, \OL{\sigma_{ij}/\sqrt\rho} +\OL{u_i\,\partial_j \sqrt\rho} \,\,\, \OL{\sigma_{ij}/\sqrt\rho}  
 +\OL{\sqrt\rho\,u_i} \,\,\, \OL{\sigma_{ij}\partial_j\left(1/\sqrt\rho\right)}\right]}_{\Sigma_\ell^{K,\mbox{\scriptsize{diss}}}}.\lb{eq:KritsukVD}~~ 
\end{eqnarray}
Here, 
\be \sigma_{ij}=2\mu(S_{ij} - \frac{1}{3} S_{kk}\delta_{ij})
\lb{eq:viscousstress}\ee 
is the deviatoric (traceless) viscous stress tensor, with the symmetric strain tensor $S_{ij} = (\partial_j u_i + \partial_i u_j)/2$. To keep the presentation simple, we assume a zero bulk viscosity even though all our analysis here and the proofs in \cite{Aluie13} (see also \cite{EyinkDrivas17a}) apply to the more general case in a straightforward manner. Superscript `F' stands for Favre, while `C' and `K'  denote the lead authors of papers in which 
those definitions, to our best knowledge, first appeared \cite{chassaing1985alternative,kida1990energy}. Terms that transport
energy conservatively in  physical space, of the form $\grad\bdot(\dots)$, are denoted by $\Sigma_\ell^{\dots,\mbox{\scriptsize{diff}}}$ for viscous diffusive transport. The remaining viscous terms are grouped together as viscous dissipative contributions and denoted by 
$\Sigma_\ell^{\dots,\mbox{\scriptsize{diss}}}$. While this grouping is the most physically sensible, we have also checked that our conclusions  hold to different terms within the grouping.
In the limit of zero filter length, {\it i.e.} in the absence of filtering, all definitions converge: 
\begin{eqnarray}
\lim_{\ell\to 0} \Sigma_\ell^F ~~~~&=& \lim_{\ell\to 0} \Sigma_\ell^C~~~= \lim_{\ell\to 0} \Sigma_\ell^K~~~=u_i\,\partial_j  \sigma_{ij}\\
\lim_{\ell\to 0} \Sigma_\ell^{F,\mbox{\scriptsize{diff}}} &=& \lim_{\ell\to 0} \Sigma_\ell^{C,\mbox{\scriptsize{diff}}}= \lim_{\ell\to 0} \Sigma_\ell^{K,\mbox{\scriptsize{diff}}} = \partial_j\left[u_i\,  \sigma_{ij}\right]\\
\lim_{\ell\to 0} \Sigma_\ell^{F,\mbox{\scriptsize{diss}}} &=& \lim_{\ell\to 0} \Sigma_\ell^{C,\mbox{\scriptsize{diss}}}= \lim_{\ell\to 0} \Sigma_\ell^{K,\mbox{\scriptsize{diss}}}=\left(\partial_j u_i\right)  \sigma_{ij}\lb{eq:unfilteredDissip}
\end{eqnarray}
It is straightforward to verify that $\Sigma_\ell^{F,\mbox{\scriptsize{diss}}}$ is Galilean invariant for any $\ell$, whereas $\Sigma_\ell^{C,\mbox{\scriptsize{diss}}}$ and $\Sigma_\ell^{K,\mbox{\scriptsize{diss}}}$ are not. Since viscous dissipation should satisfy Galilean invariance, this is one indication that the non-Favre decompositions introduce spurious effects to the large scale dynamics which are inconsistent with the physical role of viscosity. The violation of Galilean invariance would be moot if $\Sigma_\ell^{C,\mbox{\scriptsize{diss}}}$ and $\Sigma_\ell^{K,\mbox{\scriptsize{diss}}}$ were negligible, but we will show in section \ref{sec:Results} below that they are in fact quite significant. We shall now recap why the Favre scale decomposition satisfies the inviscid criterion, {\it i.e.} why $\Sigma_\ell^F(\bx)$ is guaranteed to be negligible everywhere in the domain (not just on average) at length-scales $\ell$ that are large relative to the viscous scales.

\subsection{Brief Recap of the Proof}
It was shown in \cite{Aluie13} that if 3rd-order moments of the velocity are finite, 
$\langle|\bu|^3\rangle= \frac{1}{V}\int d\bx |\bu(\bx)|^3<\infty$,
then it can be rigorously proved that $\Sigma_\ell^{F,\mbox{\scriptsize{diss}}}(\bx)$
is bounded by $O\left( \mu \,\langle|\bu|^3\rangle^{2/3}/\ell^{2} \right)$ at every point $\bx$. 
The finiteness of $\langle|\bu|^3\rangle^{1/3}$ condition is \emph{almost} as weak as requiring that the flow have finite energy and, therefore, is expected to hold in flows of interest. The type of proof used is standard in real analysis and for details pertaining 
to turbulence theory, see \cite{EyinkNotes} and Appendix A in \cite{Aluie13}.
We remark that the derivation of the bound does \emph{not} rely on the presence of turbulence. 
However, if the Reynolds number based on scale $\ell$ is small, the bound itself can be 
non-negligible since $\ell$ would not be large relative to the viscous scales.

Therefore, at high Reynolds numbers, when $\mu \,u^2_{rms}/\ell^{2} \ll 1$, KE at large `length-scales,' defined as $\OL\rho_\ell|\wt{\bu}_\ell|^2/2$ within the Favre decomposition, cannot be directly dissipated by molecular viscosity. Such KE must undergo a cascade, or an inviscid nonlinear transfer to smaller scales before it can be efficiently dissipated. A similar bound can be derived for the viscous diffusion term, $\Sigma_\ell^{F,\mbox{\scriptsize{diff}}}(\bx)$, which implies that KE at large `length-scales' does not diffuse due to molecular viscosity.

The idea behind the proof is simple and purely kinematic. A spatial derivative of a filtered field, such as $\grad \OL{f}_\ell(\bx)$, has to be bounded in magnitude by $O\left(f_{rms}/\ell\right)$. The larger is the length-scale, the smaller is the bound as one would expect. Note that the filtered gradients are bounded at every point $\bx$ in the domain. For this to hold, it is necessary to be able to commute the gradient with the filtering operation. However, a \emph{nonlinear} term such as $\OL{\left(g\grad f\right)}_\ell$, for general fields $f(\bx)$ and $g(\bx)$, cannot be expressed as a gradient of a filtered quantity and, hence, cannot be shown to be bounded. This is especially pertinent to turbulent flows, where it is well-known 
(e.g. \cite{Sreenivasan84, Sreenivasan98,Pearsonetal04}) that in the limit of large Reynolds numbers (or small viscosity $\mu\to 0$) gradients grow without bound and, as a result, a term such as $\OL{\left(g\grad f\right)}_\ell$ is expected to diverge, unless there are significant cancellations. 

For simplicity, assume for now that viscosity is spatially constant (the proofs in \cite{Aluie13} were extended to the more general case of spatially varying viscosity). It should be straightforward to verify that all derivatives  appearing in the Favre viscous terms, $\Sigma_\ell^{F,\mbox{\scriptsize{diff}}}$ and  $\Sigma_\ell^{F,\mbox{\scriptsize{diss}}}$, can be taken outside the filtering operation. It follows that each of $\Sigma_\ell^{F,\mbox{\scriptsize{diff}}}(\bx)$ and  $\Sigma_\ell^{F,\mbox{\scriptsize{diss}}}(\bx)$ can be rigorously bounded by $O\left( \mu/\ell^{2} \right)$, which vanishes in high Reynolds number flows when $\ell$ is large compared with the viscous cut-off scale \cite{Aluie13}. The situation is different for the other two decompositions. For example, the term $\OL{\left(\sigma_{ij}/\rho\right)}_\ell$ appearing in $\Sigma_\ell^{C}$ $\left(\mbox{eq.} \left(\ref{eq:ChassaingVD}\right)\right)$ is similar to $\OL{\left(g\grad f\right)}_\ell$ and cannot be rewritten as a gradient of a filtered quantity and, hence, cannot be bounded in the presence of significant density variations. While we are unable to prove mathematically that viscous terms, $\Sigma_\ell^{C}$ and $\Sigma_\ell^{K}$, do not vanish when $\ell$ is large, we shall now present numerical evidence that such is the case. From a mathematical standpoint, these different decompositions correspond to different ways to regularizing the equations as was highlighted recently by Eyink \& Drivas \cite{EyinkDrivas17a}. They used the inviscid criterion to 
extend the coarse-graining analysis to internal energy and analyzed the inertial-range dynamics for what they called ``intrinsic large-scale internal energy.''

In the next section, we test if a scale decomposition satisfies the inviscid criterion by fixing viscosity, $\mu$,
and analyzing the viscous contributions as a function of length-scale $\ell$. This allows us to use a single 
simulation for each of our tests. 
Another way to carry out such tests is by analyzing the viscous contributions at a fixed scale $\ell$
while varying $\mu$. This second way is equivalent to the first in the sense that $\ell$ is made 
`larger' relative to the viscous scale by taking $\mu\to0$ rather than $\ell\to\infty$ as in the first approach.
While taking the limit $\mu\to 0$ is still of theoretical and practical interest, it is computationally quite expensive
since it requires a series of simulations with a progressively smaller viscosity for every single test.

\section{Numerical Results\lb{sec:Results}}
In this section, we shall present numerical results from flows of a 1D shock, and the Rayleigh-Taylor Instability in 2D and 3D.
We use the fully compressible Navier-Stokes equations
\begin{eqnarray} 
&\hspace{-0.4cm}\partial_t \rho& + \partial_j(\rho u_j) = 0 \lb{continuity} \\
&\hspace{-0.4cm}\partial_t (\rho u_i)& + \partial_j(\rho u_i u_j) 
= -\partial_i P +  \partial_j\sigma_{ij} - \rho \,g \,\delta_{zi}   \lb{momentum}\\
&\hspace{-0.4cm}\partial_t (\rho E)&+ \partial_j(\rho E u_j) 
= -\partial_j (P u_j) +\partial_j[2\mu ~ u_i(S_{ij} - \frac{1}{d} S_{kk}\delta_{ij})]  -\partial_j q_j -\rho u_i  \,g \,\delta_{zi}  \lb{total-energy}\end{eqnarray}
Here, $\bu$ is velocity, $\rho$ is density, $E=|\bu|^2/2 + e$ is total energy per unit mass, where $e$ is specific internal energy,  $P$ is thermodynamic pressure, $\mu$ is dynamic viscosity, ${\bf g}$ is gravitational acceleration along the vertical z-direction, 
${\bf q} = -\kappa \grad T$ is the heat flux with a thermal conductivity $\kappa$ and temperature $T$. We use the ideal gas equation of state (EOS).
$S_{ij}$ is the symmetric strain tensor and $\sigma_{ij}$ is the viscous stress defined in eq. (\ref{eq:viscousstress}). 
In the flows we analyzed, we considered both spatially constant and spatially varying dynamic viscosity and thermal conductivity, as we  elaborate below. However, we found that our results are insensitive to this choice.

When calculating viscous terms in eqs. (\ref{eq:FavreVD})-(\ref{eq:KritsukVD}), the fields are filtered using a Gaussian kernel,
\be G_\ell({|\bx|})=\left(\frac{6}{\pi \ell^2}\right)^{n/2}e^{-\frac{6}{\ell^2}|\bx|^2}.
\ee
in dimensions $n=1,2,3$. This Gaussian kernel form has been used in several prior studies (e.g. \cite{Piomellietal91,Wangetal18}) due to advantages  in numerical discretization (see \cite{john2012large}, page 30). In this work, we purposefully avoid using a sharp-spectral filter which, for density, yields $\OL\rho_\ell(\bx)$, with Fourier modes larger than $\gtrsim \ell^{-1}$ discontinuously truncated. Such coarse-graining of density violates physical realizability since it can have negative values due to the non-positivity of the sharp-spectral filter in x-space \cite{Aluie13}.
In our RT flows, which have no-slip rigid walls at the top and bottom boundaries, filtering near the walls is performed by extending the computational domain in accordance with the boundary conditions. To be specific, beyond the wall, the density field is kept constant (zero normal gradient) and the velocity is kept zero.

\subsection{1D Normal Shock}
We first test our hypothesis in a simple 1-dimensional steady shock solution of eqs. (\ref{continuity})-(\ref{total-energy}). Here we shall show that unlike the Favre decomposition, the alternate two decompositions yield a significant viscous contribution at large `length-scales'  at a moderate transonic Mach number,  which becomes even more pronounced in a Mach 3 shock.

\begin{figure*}[bhp]
\vspace{-.5cm} 
\begin{minipage}[b]{1.0\textwidth}  
\subfigure[\footnotesize{Density}]{\includegraphics[height=2.6in]{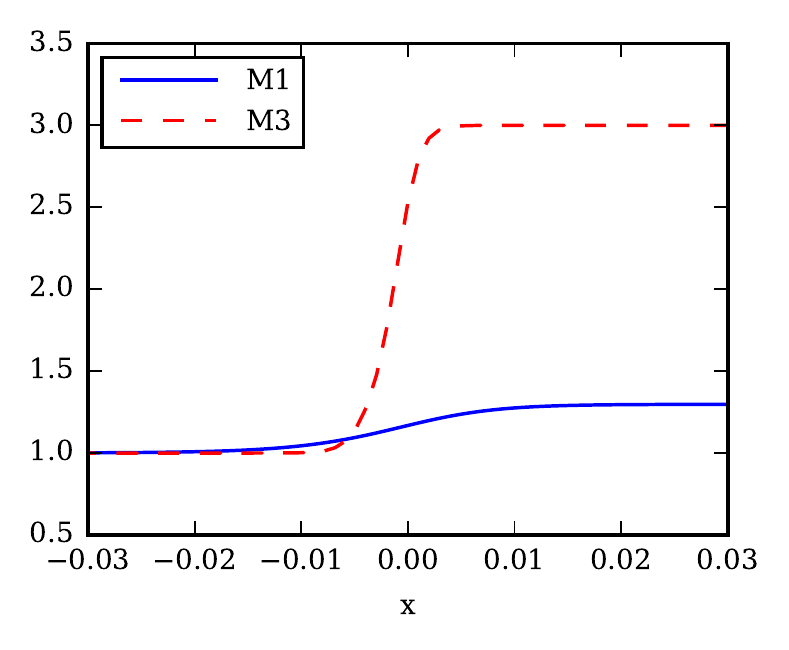}}
\phantom{}
\subfigure[\footnotesize{Velocity}]{\includegraphics[height=2.6in]{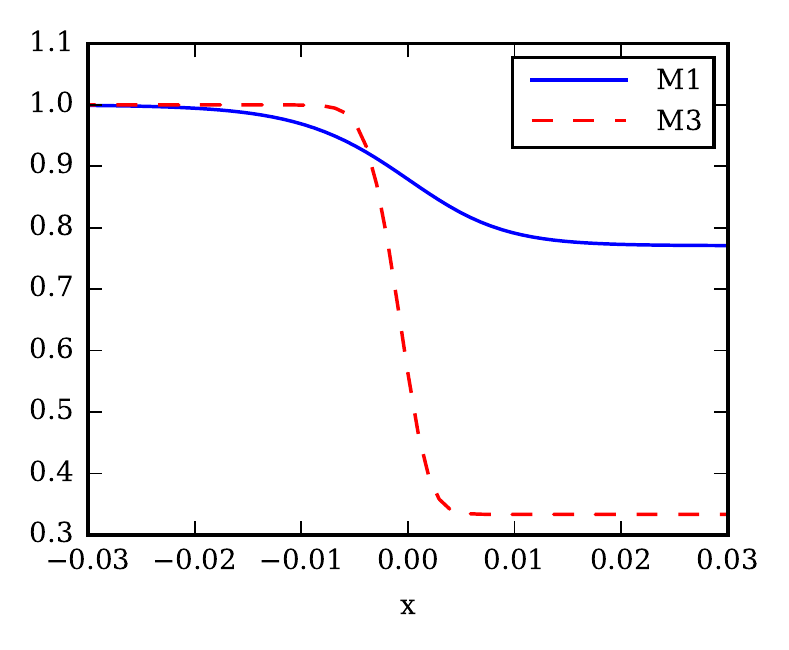}}
\caption{\footnotesize{Shock profiles at Mach 1.2 (M1) and 3 (M3) in the reference frame of the shock. } \label{fig:Viz_1Dshock}}
\end{minipage}
\end{figure*}

Equations (\ref{continuity})-(\ref{total-energy}) with zero gravity are solved numerically starting from the Rankine-Hugoniot jump conditions. The solutions are in the shock frame of reference and are shown in Fig. \ref{fig:Viz_1Dshock}. The parameters we consider are in Table \ref{tbl:Params_1Dshock}. Zero-gradient boundary conditions (BC) apply at the boundaries of the our domain. We use subscript `$0$' for the upstream/pre-shock region and `$1$' for the post-shock/downstream region. In addition to the two cases in Table \ref{tbl:Params_1Dshock}, we also analyzed a Mach 3 shock with constant viscosity. The results (see Appendix) are very similar to those presented here at the same Mach number, indicating that a variable viscosity does not affect our conclusions.
It is perhaps worth noting that in the 1D shock context, $\mu$ in the viscous stress, eq. (\ref{eq:viscousstress}), can be regarded as the sum of dynamic viscosity and $3/4$-th times the bulk viscosity (e.g. \cite{Johnson13}). 

Equations of conservation of mass, momentum, and energy fully determine the post-shock flow variables, $\rho_1$, $u_1$, and $p_1$, from their pre-shock counterparts, $\rho_0$, $u_0$, and $p_0$: 
\begin{eqnarray}
m_0 &=& \rho_0 u_0 = \rho_1 u_1\nonumber\\
m_0V_0 &=& \rho_0 u^2_0 + p_0= \rho_1 u^2_1+ p_1 \\
m_0I_0 &=& \left(\frac{1}{2}\rho_0 u^2_0 + \frac{\gamma}{\gamma-1}p_0\right)u_0= \left(\frac{1}{2}\rho_1 u^2_1 + \frac{\gamma}{\gamma-1}p_1\right)u_1 \nonumber
\end{eqnarray}
The solution can be normalized by the three dynamical invariants, $m_0$, $m_0V_0$, and $m_0I_0$,  which are three independent parameters set as boundary conditions (e.g. \cite{ZeldovichRaizer02}). Fixing a pre-shock Mach number, $M_0$, is equivalent to fixing the ratio of ram pressure to thermodynamic pressure, $\rho_0u_0^2/\gamma p_0$, which effectively fixes $I_0$, leaving two free parameters, $m_0$ and $V_0$. In what follows, we shall normalize our results in terms of $\rho_0$ and $u_0$.

There are two length-scales of interest to us in this problem. The viscous scale of shock, \be
\ell_\mu\equiv \frac{\mu_0}{\rho_0\left(u_0-u_1\right)} = \frac{\mu_0}{\rho_0 u_0}\left(\frac{\gamma+1}{2}\frac{M^2_0}{M^2_0-1}\right),
\lb{eq:shock_width}\ee
and a characteristic macroscopic length-scale determined 
from the Reynolds number (which is arbitrary in this simple shock solution) and is independent of the Mach number,
\be
L \equiv  \frac{\mu_0}{\rho_0 u_0} Re_0~.
\lb{eq:shock_LargeScale}\ee
Their ratio is solely a function of the Reynolds and Mach numbers:
\be \frac{L}{\ell_\mu} = Re_0 \left(1-\frac{1}{M^2_0}\right)\frac{2}{\gamma+1}.
\lb{eq:shock_scaleRange}\ee
In what follows, we shall define length-scale $\ell$ in relation to the macroscopic scale $L$ due to its independence of $M_0$. This allows us to compare scales in flows at the same $Re_0$ but at different $M_0$.

\begin{figure*}[!htb]
\vspace{-0.7cm}
\begin{minipage}[b]{1.0\textwidth}  
\subfigure[\footnotesize{Viscous contribution at L/8}]
{\includegraphics[height=2.6in]{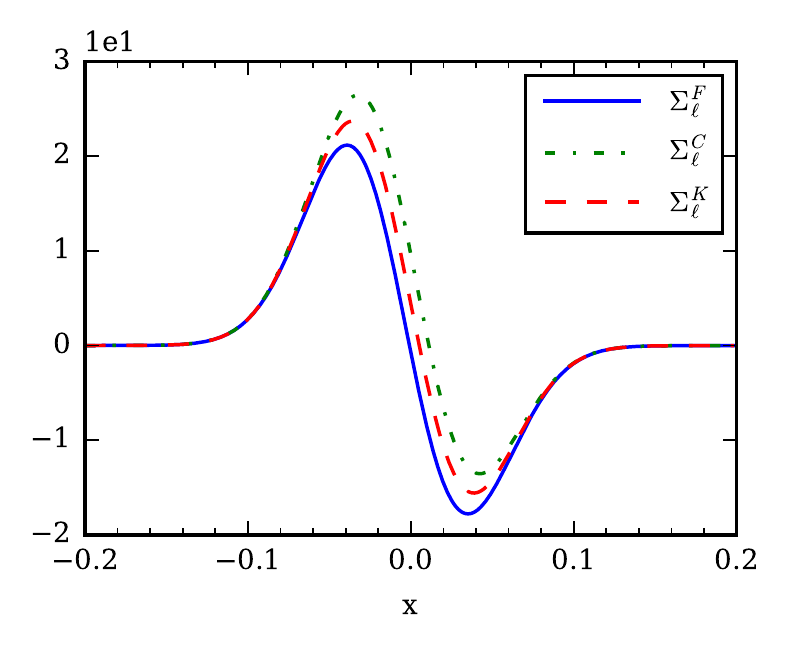}} 
\phantom{}
\subfigure[\footnotesize{Dissipation at L/8}]
{\includegraphics[height=2.6in]{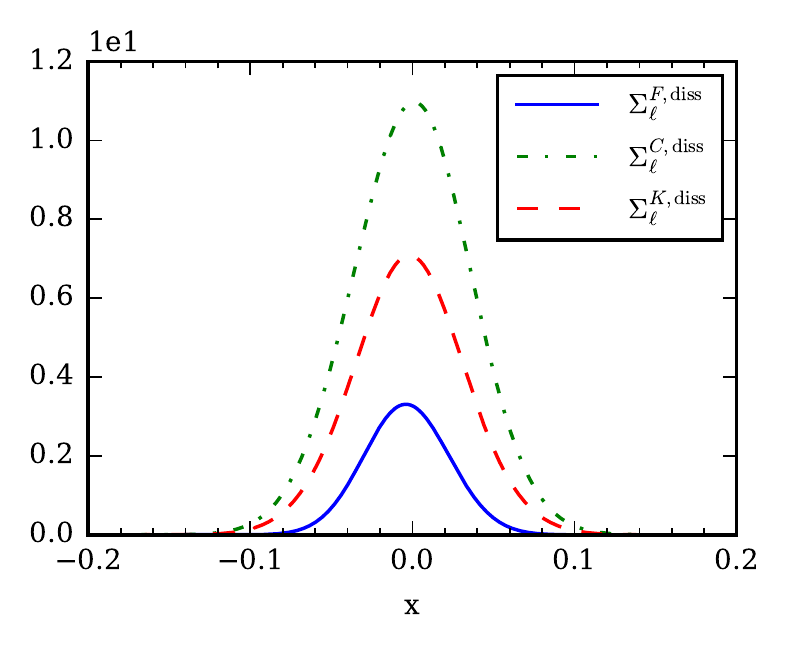}}  \\
\phantom{}
\subfigure[\footnotesize{Viscous contribution at L/8}]
{\includegraphics[height=2.6in]{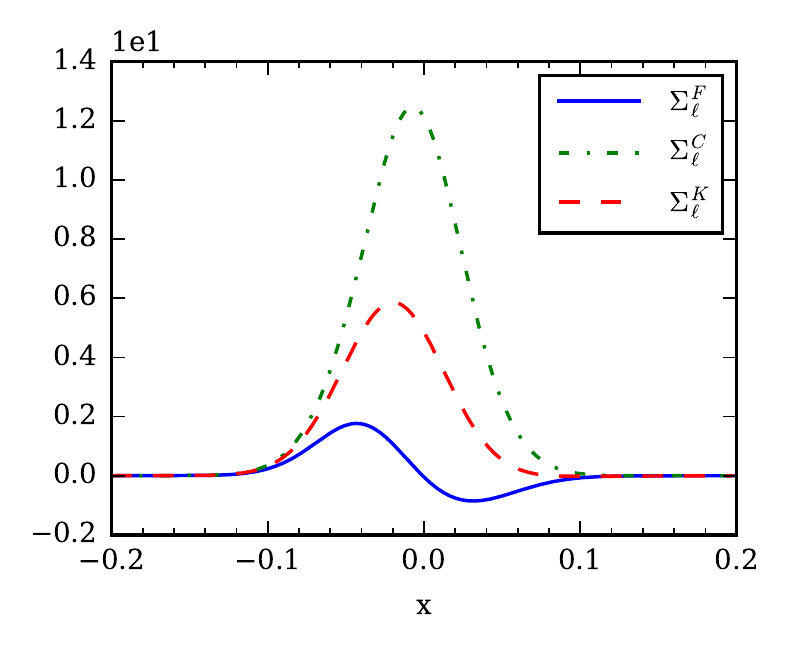}} 
\phantom{}
\subfigure[\footnotesize{Dissipation at L/8}]
{\includegraphics[height=2.6in]{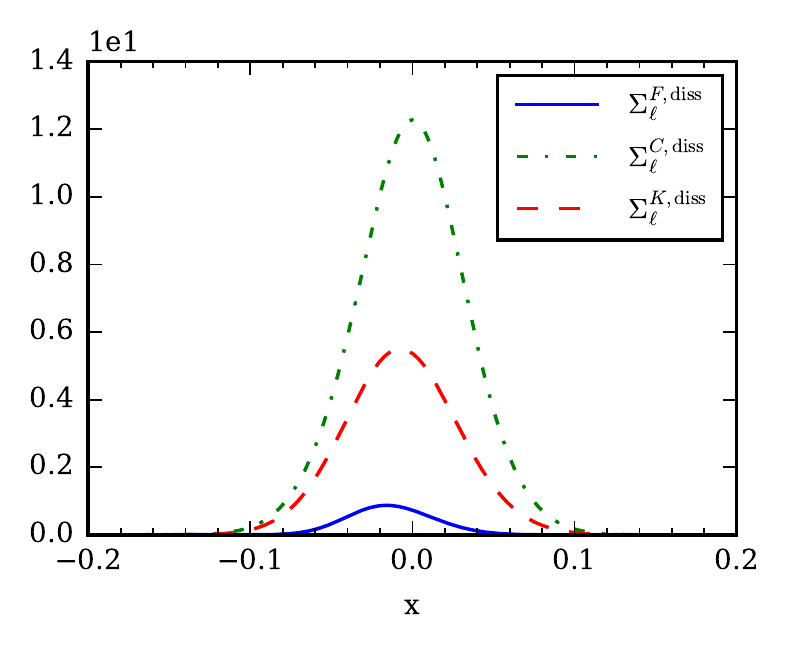}}
\caption{\footnotesize{Mach 1.2 (top row) and  3 (bottom row). The viscous contributions are calculated at `length-scale' $L/8$. The data is normalized by the unfiltered dissipation. The Favre decomposition yields a smaller viscous contribution at large `length-scales' compared to the other two decompositions. Moreover, the disparity between decompositions increases with Mach number.}\label{fig:Shock_dissip_L8}}
\end{minipage}
\end{figure*}

The dissipation terms, $\Sigma_\ell(\bx)$ and $\Sigma_\ell^{\mbox{\scriptsize{diss}}}(\bx)$, using the three decompositions at length-scale $\ell=L/8$, are plotted as a function of $\bx$ in Fig. \ref{fig:Shock_dissip_L8}. It shows that at both Mach numbers, the Favre decomposition yields the smallest viscous contribution to the `large-scale' dynamics. We also observe that the discrepancy between the three decompositions increases with higher $M_0$. As we have discussed, in the limit of zero density gradients, all three decompositions converge, while in the limit of high Mach numbers and increasing density differences, the discrepancy between the three decompositions is expected to grow. Notice that $\Sigma_\ell^{\mbox{\scriptsize{F,diss}}}(\bx)$ and $\Sigma_\ell^{\mbox{\scriptsize{K,diss}}}(\bx)$ are both asymmetric around the shock, which is due to the density-weighting.

\begin{figure*}[htb] 
\begin{minipage}[b]{1.0\textwidth}  
\subfigure[\footnotesize{Dissipation average}]
{\includegraphics[height=2.5in]{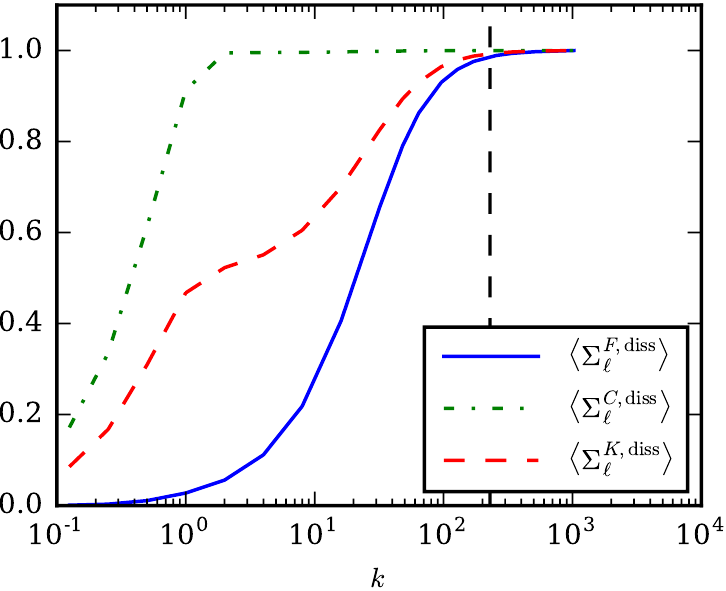}} 
\phantom{}
\subfigure[\footnotesize{Dissipation $L^\infty$ log-log}]{\includegraphics[height=2.5in]{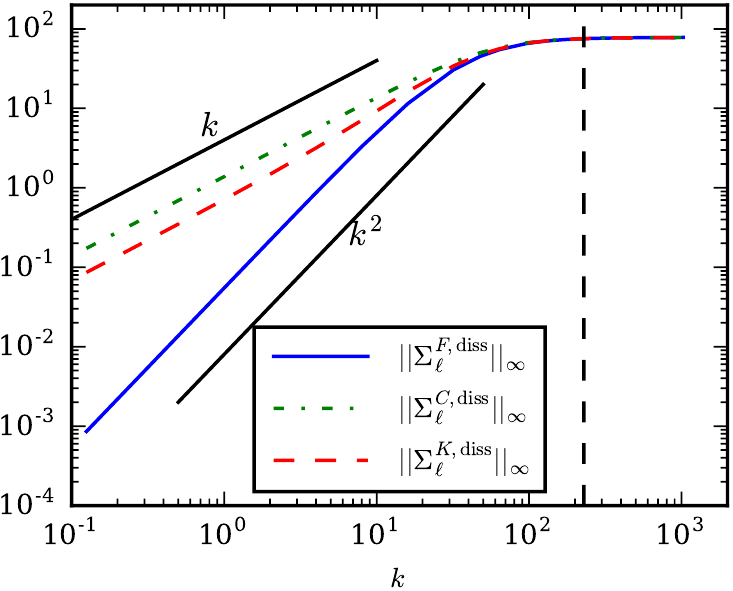}}\\
\phantom{}
\subfigure[\footnotesize{Dissipation average}]
{\includegraphics[height=2.5in]{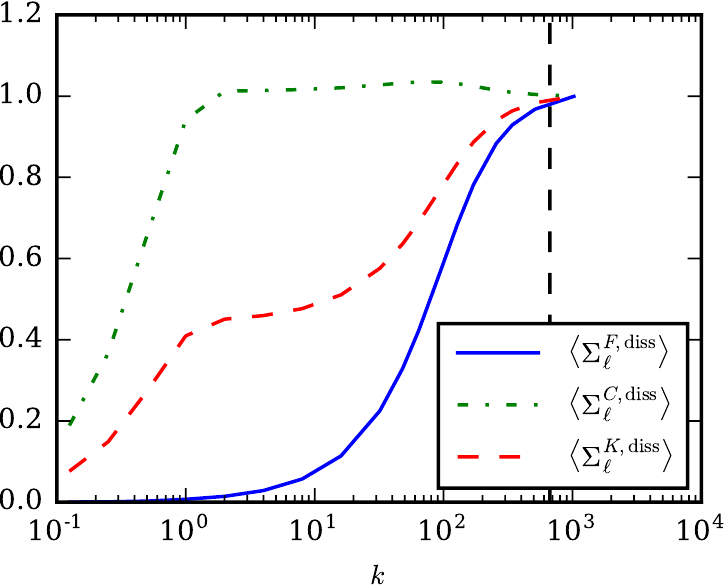}} 
\phantom{}
\subfigure[\footnotesize{Dissipation $L^\infty$ log-log}]{\includegraphics[height=2.5in]{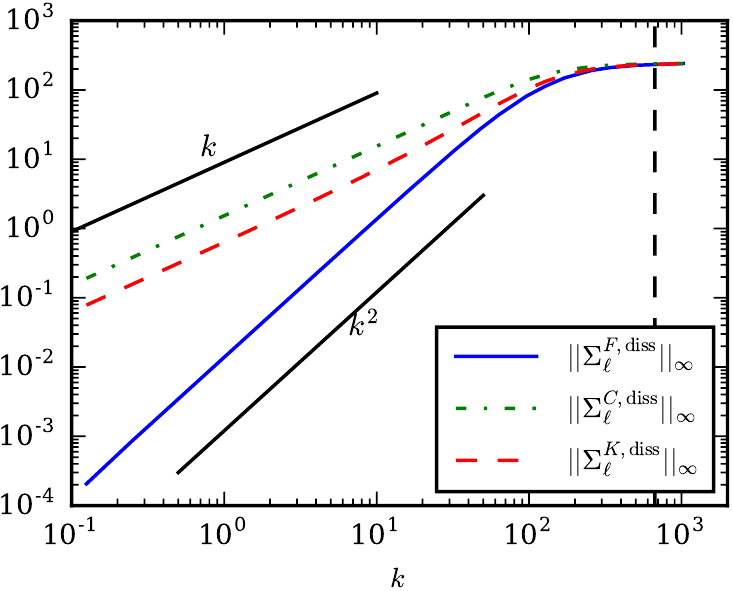}}  
\caption{\footnotesize{Mach 1.2 (top row) and  3 (bottom row). The figure shows both the average and the maximum dissipation as a function of $k=L/\ell$, and the vertical line in each figure marks the viscous cut-off wavenumber $k_d=k_\mu=L/\ell_\mu$. Data is normalized by the unfiltered dissipation. Left two panels show how the non-Favre decompositions yield significant viscous contamination at large `length-scales.' Right two panels show that $\Sigma^{F,\text{diss}}$ decays as $\ell^{-2}$ as proven mathematically whereas the non-Favre definitions decay at a much slower rate of $\ell^{-1}$ due to the dilution of the shock's effect in 1-dimension.
} \label{fig:Ma1.2_dissip_stat}}
\end{minipage}
\end{figure*}

The viscous dissipation as a function of `length-scale' in  plotted Fig. \ref{fig:Ma1.2_dissip_stat}.  Here, we define wavenumber as $k=L/\ell$, such that the wavenumber associated with the shock width is $k_\mu = L/\ell_\mu$ from eq. (\ref{eq:shock_scaleRange}). The left two panels in Fig. \ref{fig:Ma1.2_dissip_stat} show evidence of significant viscous contamination at intermediate to large `length-scales' within the non-Favre decompositions. The contamination also seems to increase with Mach number. This presents evidence that the two non-Favre decompositions we consider here violate the inviscid criterion and that they are not suitable to analyze inertial-range dynamics in compressible flows. The right two panels show the scaling of the $L^\infty$-norm of $\Sigma_\ell^{\mbox{\scriptsize{diss}}}(\bx)$ for the three decompositions. It shows that $\|\Sigma_\ell^{\mbox{\scriptsize{F,diss}}}\|_\infty$ varies as $\ell^{-2}$ at both Mach numbers and also for constant and spatially varying $\mu$, which is consistent with the proof of \cite{Aluie13}. This is because $\|\Sigma_\ell^{\mbox{\scriptsize{F,diss}}}\|_\infty$ is the upper bound of the pointwise quantity $|\Sigma_\ell^{\mbox{\scriptsize{F,diss}}} (\bx)|$ and, therefore, the dissipation has to vanish at every point $\bx$ at least as fast as $\ell^{-2}$ for large $\ell$. 

On the other hand, the non-Favre dissipation terms vary as a $\ell^{-1}$. While this is a weaker decay rate than that obtained by a Favre decomposition, it suggests that viscous contributions perhaps do vanish in the limit of large length-scales. However, the $\ell^{-1}$ decay is due to the presence of just one singular structure (the shock) whose effect is diluted by filtering over an ever-wider domain in 1 dimension. We will present evidence below that this trend does not hold in more complex flows.

\begin{table} \centering
\setlength{\tabcolsep}{4pt}
\caption{Parameters of the 1D shock solutions shown in Fig. \ref{fig:Viz_1Dshock}. Unlike run M3, M1 uses a spatially constant dynamic viscosity $\mu_0$. Both solutions are for a Prandtl number of unity, $Pr=\mu C_p/\kappa =1$. The shock width $\ell_s$, which is associated with $\ell_\mu$, is calculated from the maximum gradient scale length $\ell_s=\text{max}\frac{|u|}{|\nabla u|}$. Unfiltered dissipation $\langle\Sigma^{\mbox{\scriptsize{diss}}}_{\ell=0}\rangle$ is given by eq. (\ref{eq:unfilteredDissip}).} 
\begin{tabular}{|c|c |c|c|c|c|c|c|c|c|}
\hline  & $Ma$ & $Re_0$ & $\mu_0$ & $\mu(\bx)$ &$Pr$ &$L$ &$\ell_s$ &$\ell_\mu$ & $\langle\Sigma^{\mbox{\scriptsize{diss}}}_{\ell=0}\rangle$ \\ \hline
M1 & $1.2$  & 1000 &$10^{-3}$ & $\mu_0$ & 1 &$1$ &$4.39\times 10^{-2}$ & $4.36\times 10^{-3}$ & $1.82\times 10^{-3}$  \\ \hline
M3 & $3$  & 1000 &$10^{-3}$ & $\mu_0T^{0.76}$ & 1 &$1$ &$1.07\times 10^{-2}$ & $1.50\times 10^{-3}$ & $3.03\times 10^{-2}$\\ \hline
\end{tabular}
\label{tbl:Params_1Dshock}
\end{table}

\clearpage
\newpage

\subsection{2D Rayleigh-Taylor Instability\lb{sec:2DRTI}}

\begin{figure*}[htb]
\begin{minipage}[b]{1.0\textwidth}  
\subfigure[\footnotesize{R2}]
{\includegraphics[height=2.2in]{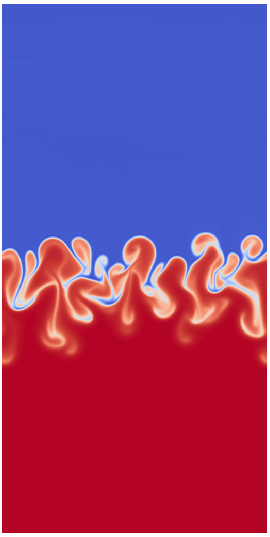}} 
\phantom{aaa}
\subfigure[\footnotesize{R3}]
{\includegraphics[height=2.2in]{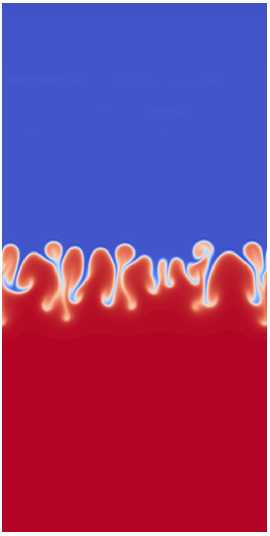}} 
\phantom{aaa}
\subfigure[\footnotesize{R4}]
{\includegraphics[height=2.2in]{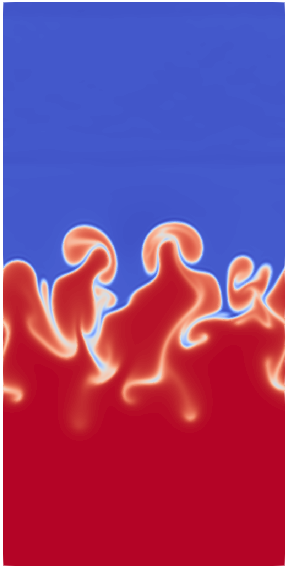}} 
\\
\subfigure[\footnotesize{R4vv}]
{\includegraphics[height=2.2in]{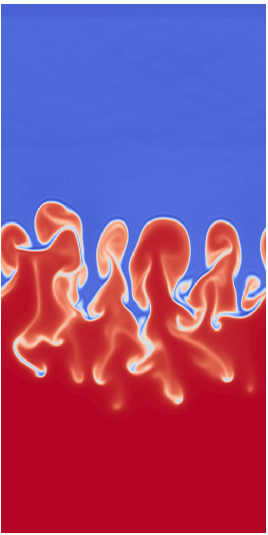}} 
\phantom{aaa}
\subfigure[\footnotesize{R4vc}]
{\includegraphics[height=2.2in]{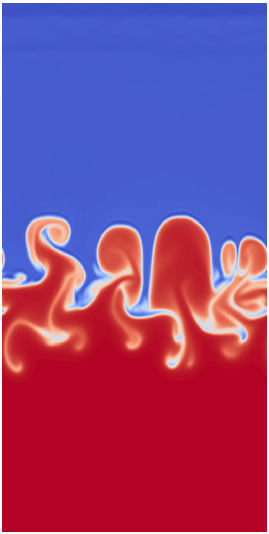}} 
\caption{\footnotesize{Density field of 2D buoyancy-driven flows R2-R4 carried out with successively higher initial density ratios (see Table \ref{tbl:Params_2DRTI}). Flows R4vv and R4vc in the bottom two panels test the sensitivity of our results to the temperature-dependence of viscosity, $\mu$,  and thermal conductivity coefficient, $\kappa$.} \label{fig:Viz_2DRTI}}
\end{minipage}
\end{figure*}

\begin{table} \centering
\setlength{\tabcolsep}{4pt}
\caption{Parameters of the flows shown in Fig.~\ref{fig:Viz_2DRTI}.  Prandtl number is $Pr_0=C_p\mu_0/\kappa_0$, mesh Grashof number is $Gr=2Ag\langle \rho\rangle^2 \Delta x^3/\mu_0^2$, where $\Delta x$ the grid cell size, $g$ is gravitational acceleration, $A=(\rho_h-\rho_l)/(\rho_h+\rho_l)$ is the Atwood number, and $\mu_0$ is the average dynamic viscosity. The perturbation Reynolds number is $Re_p=\langle \rho\rangle \lambda \sqrt{Ag\lambda/(1+A)}/\mu_0$, with $\lambda$ the largest initial perturbation wavelength. The Kolmogorov scale is $\eta=\mu_0^{3/4}/(\epsilon^{1/4}\langle \rho \rangle^{3/4})$, where $\epsilon$ is the specific kinetic energy dissipation rate. Gravitational acceleration $g$ is set to  1 in all cases. The temperature range in cases R4vv and R4vc is $T(\bx) \in [2.5, 45000]$. }
\begin{tabular}{|c|c |c|c|c|c|c|c|c|c|c|}
\hline  &$\rho_h/\rho_l$ & $\langle\rho\rangle$ & $\rho'_{rms}$ & $\rho'_{rms}/\langle \rho\rangle$ & 
 $\mu(\bx)$ & $\kappa(\bx)$ & $Pr_0$ & $Gr$ & $Re_p$ &$\eta/\Delta x$ \\ \hline
 R2 &$10^2$ & 0.501 & 0.690 & 1.377 & $6\times 10^{-5}$ & $1.2\times 10^{-4}$&  1 & 4.25 & 528 & 1.276\\ \hline
R3 & $10^3$ & 0.489 & 0.682 & 1.395 & $9\times 10^{-5}$ & $1.8\times 10^{-4}$  & 1 & 1.80& 344 & 2.031 \\ \hline
R4 & $10^4$ & 0.482 & 0.682 & 1.395 & $9\times 10^{-5}$ & $1.8\times 10^{-4}$  & 1 & 1.75 & 339 & 2.029 \\ \hline
R4vv  & $10^4$  &0.463 & 0.641 & 1.384 & $ 10^{-4}T^{0.3}$ & $2\times 10^{-4} \,T^{0.3}$  & 1 & 1.31 &293 & 1.517\\ \hline
R4vc & $10^4$  & 0.443 & 0.615 & 1.388 & $ 10^{-4}T^{0.3}$ & $2\times 10^{-4}$  & 1 &1.20 & 280 & 1.713 \\ \hline
\end{tabular}
\label{tbl:Params_2DRTI}
\end{table}

Equations (\ref{continuity})-(\ref{total-energy}) with $g=1$  are used to conduct five different simulations of the Rayleigh-Taylor instability (RTI) in 2D using our code DiNuSUR. We impose no-slip BC at the top and bottom walls and periodic BC in the horizontal direction.  All five runs were carried out on a  $N_x\times N_z= 512\times 1{,}024$ grid using a pseudospectral solver in the horizontal direction and a 6th-order compact finite difference scheme in the z-direction. The physical dimensions of the domain are $L_x\times L_z=  1.6\times 3.2$. The initial conditions of the simulations are those of a heavy fluid with $\rho_h=1.0$, filling the top-half of the domain in the z-direction, over a lighter fluid with density $\rho_l$ in the bottom half. The initial pressure satisfies the hydrostatic equilibrium $dP/dz=-\rho g$, and the initial velocity is zero with velocity perturbation added at the interface. Small amplitude perturbations result in RTI which evolves until the times shown in Fig.~\ref{fig:Viz_2DRTI}, which are the snapshots we analyze. The specific time at which we analyze the flow is not special except in that the flow has to develop sufficiently for the nonlinearities to become significant. We have checked that our conclusions hold at other times (see Appendix). The snapshots from the five flows we analyze are highly nonlinear (the density modulation amplitude exceeds the perturbation wavelength) but not fully turbulent. The Kolmogorov dissipative length scale in Table \ref{tbl:Params_2DRTI} is larger than the grid cell size in all our cases. The Grashof number is slightly larger than unity, which indicates that our simulations may become under-resolved at much later times when the flow becomes turbulent \cite{WeiLivescu12}.

Dynamic viscosity, $\mu$, in some of our simulations was taken to be spatially constant similar to previous studies of Rayleigh-Taylor turbulence \cite{LivescuRistorcelli,CabotCook}. In two of our simulations, we  also used a temperature-dependent viscosity, $\mu(\bx)=\mu_0 (T(\bx)/T_0)^{\alpha}$, with $\alpha=0.3$. We have taken $\alpha$ smaller than the usual $\alpha=0.76$, which was computationally too expensive to numerically solve the equations with hot-to-cold temperature ratios that are several orders of magnitude large.

\begin{figure*}[bhp]
\begin{minipage}[b]{1.0\textwidth}  
\subfigure[\footnotesize{R2: $\langle\Sigma_\ell\rangle$}]
{\includegraphics[height=1.68in]{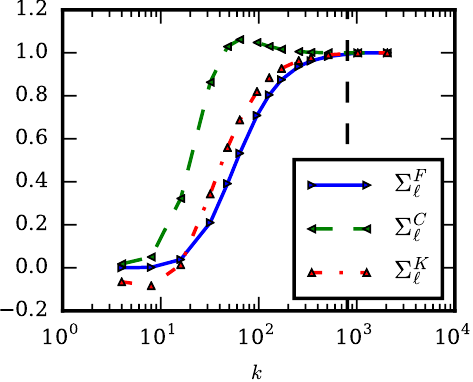}} 
\phantom{}
\subfigure[\footnotesize{R3: $\langle\Sigma_\ell\rangle$}]
{\includegraphics[height=1.68in]{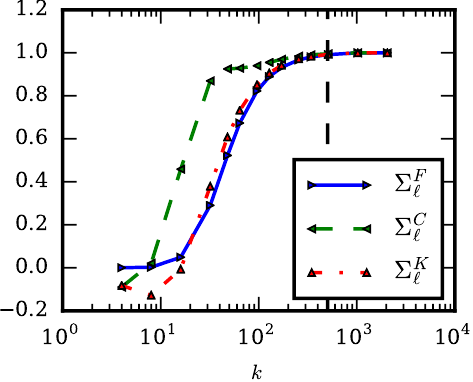}  \label{fig:R3_ave_early}} 
\phantom{}
\subfigure[\footnotesize{R4: $\langle\Sigma_\ell\rangle$}]
{\includegraphics[height=1.68in]{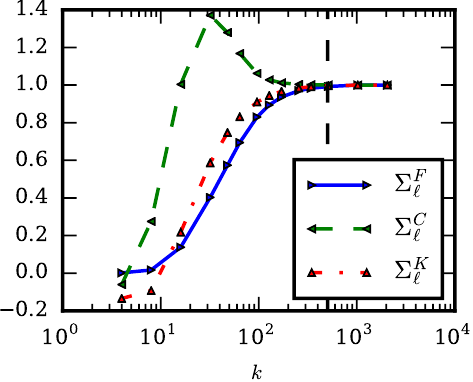}} 
\\
\subfigure[\footnotesize{R2: $\|\Sigma^{\mbox{\scriptsize{diss}}}_\ell\|_1$}]{\includegraphics[height=1.68in]{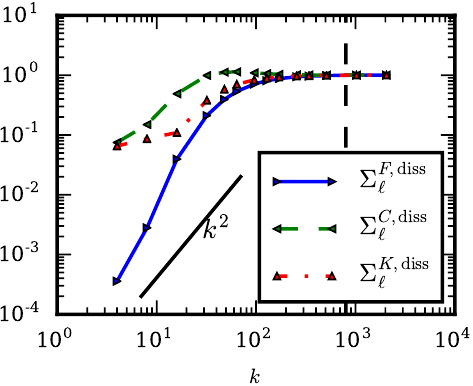}}
\phantom{}
\subfigure[\footnotesize{R3: $\|\Sigma^{\mbox{\scriptsize{diss}}}_\ell\|_1$}]{\includegraphics[height=1.68in]{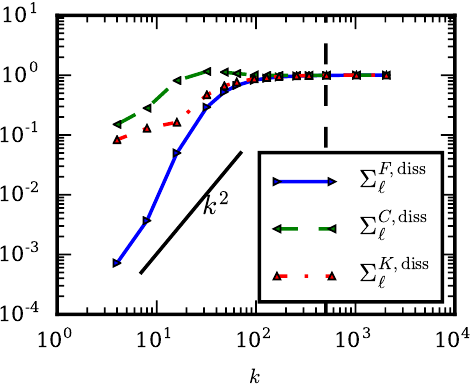} \label{fig:R3_l1_early}}
\phantom{}
\subfigure[\footnotesize{R4: $\|\Sigma^{\mbox{\scriptsize{diss}}}_\ell\|_1$}]{\includegraphics[height=1.68in]{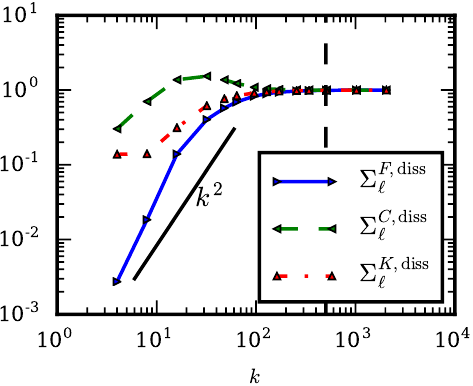}}
\caption{\footnotesize{Comparing the viscous contribution from the decompositions as a function of scale using three flows R2-R4 with increasing density ratio from left to right (see Table \ref{tbl:Params_2DRTI}). Horizontal axes are 
$k=L_z/\ell$, and the vertical line in each figure marks the viscous cut-off wavenumber $k_d=L_z/\eta$. Top row shows the full viscous contribution $\langle\Sigma_\ell\rangle$.
Bottom row shows the $L^1$-norm scaling of the viscous dissipation, $\|\Sigma^{\mbox{\scriptsize{diss}}}_\ell\|_1$. Plots are normalized by the unfiltered dissipation. In all cases, the Favre dissipation decays at least as $\ell^{-2}$ at large scales. The other two definitions do not show a clear decay trend and even become negative on average at the largest scales.}\label{fig:Decomps_cstMu_2DRTI}}
\end{minipage}
\end{figure*}

Fig. \ref{fig:Decomps_cstMu_2DRTI} measures the average viscous contribution $\langle\Sigma_\ell\rangle$, which includes dissipation and diffusion effects in flows with increasing density ratios. In these RT flows with zero in/out flow boundary conditions, the contribution from diffusive terms, $\langle\Sigma_\ell^{\mbox{\scriptsize{diff}}}\rangle$ in eqs. (\ref{eq:FavreVD})-(\ref{eq:KritsukVD}),  is negligibly small (by a factor $10^{-6}$ or smaller relative to dissipation) on average at all length-scales we analyzed. In such complex flows, the $L^\infty$-norm is not a robust metric, unlike in the 1D shock problem of the previous subsection. To gauge the pointwise dissipation and in order to avoid cancellations from the spatial averaging ($\Sigma^{\mbox{\scriptsize{diss}}}_\ell(\bx)$ is not positive definite), we use the $L^1$-norm as a metric: $\|\Sigma^{\mbox{\scriptsize{diss}}}_\ell\|_1 = \langle|\Sigma^{\mbox{\scriptsize{diss}}}_\ell |\rangle$.

In the large density ratio simulations, R3 and R4, the flows do not become very turbulent in the course of their development. The viscous terms in all three cases in Fig. \ref{fig:Decomps_cstMu_2DRTI} exhibit a similar trend with length-scale, despite the higher density ratios in R3 and R4.
We induce that density variations alone are not sufficient to yield large differences between the three decompositions, but that velocity fluctuations (or velocity gradients) are just as important. 

Nevertheless, we still observe marked differences between the three decompositions: (i) $\langle\Sigma^{\mbox{\scriptsize{C}}}_\ell\rangle$ is significant and contaminates a wider range of scale before it decays. Moreover, in cases R3 and R4, it becomes negative and grows in magnitude again at the largest scales. (ii) While $\langle\Sigma^{\mbox{\scriptsize{K}}}_\ell\rangle$ is fairly close in value to its Favre counterpart over a range of $\ell$, it diverges from it and becomes negative (growing in magnitude) at the largest scales in all three cases R2-R4. (iii) The clearest distinction between the three quantities is seen when considering $\|\Sigma^{\mbox{\scriptsize{diss}}}_\ell\|_1$ as a proxy for the pointwise behavior of viscous dissipation. While $\|\Sigma^{\mbox{\scriptsize{F,diss}}}_\ell\|_1$ decays at least as fast as $\ell^{-2}$ at large scales, the other two definitions do not show a clear decay trend for decay and are several orders of magnitude larger than $\|\Sigma^{\mbox{\scriptsize{F,diss}}}_\ell\|_1$, precluding inertial dynamics at those `length-scales.'

\begin{figure*}[bhp]
\begin{minipage}[b]{1.0\textwidth}  
\subfigure[\footnotesize{R4vv: $\langle\Sigma_\ell\rangle$}]
{\includegraphics[height=1.73in]{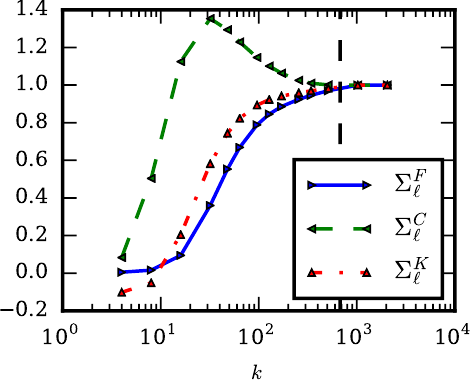}} 
\phantom{}
\subfigure[\footnotesize{R4vc: $\langle\Sigma_\ell\rangle$}]
{\includegraphics[height=1.73in]{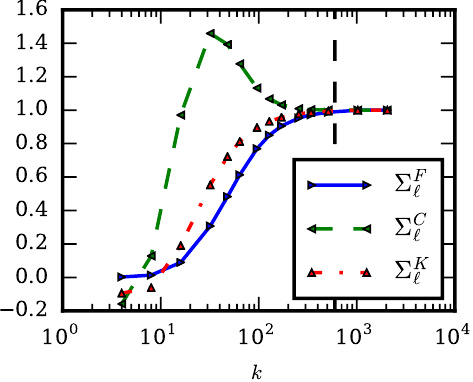}} 
\\
\subfigure[\footnotesize{R4vv: $\|\Sigma^{\mbox{\scriptsize{diss}}}_\ell\|_1$}]{\includegraphics[height=1.68in]{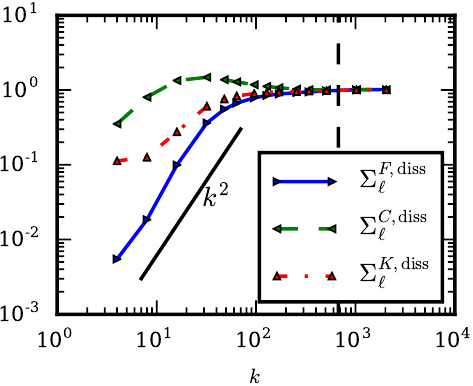}}
\phantom{}
\subfigure[\footnotesize{R4vc: $\|\Sigma^{\mbox{\scriptsize{diss}}}_\ell\|_1$}]{\includegraphics[height=1.68in]{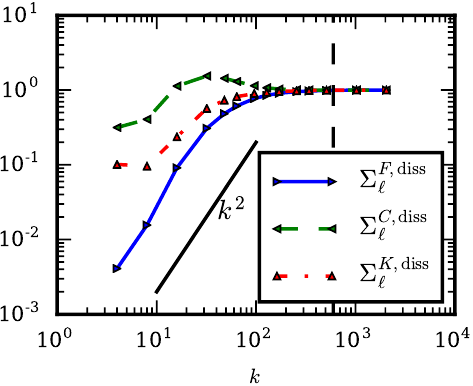}}

\caption{\footnotesize{Similar to Fig.~\ref{fig:Decomps_cstMu_2DRTI}. Testing the sensitivity of results to a spatially varying viscosity and thermal conductivity (runs R4vv and R4vc in Table \ref{tbl:Params_2DRTI}). Plots are normalized by the unfiltered dissipation. The plots are consistent with those in Fig. \ref{fig:Decomps_cstMu_2DRTI}, showing that our conclusions also hold when $\mu(\bx)$ and $\kappa(\bx)$ are spatially varying, as proved in \cite{Aluie13}.}\label{fig:Decomps_varMu_2DRTI}}
\end{minipage}
\end{figure*}

While the three cases R2-R4 were carried out with a constant viscosity and thermal conductivity, Fig. \ref{fig:Decomps_varMu_2DRTI} tests the sensitivity of our results to spatially varying $\mu(\bx)$ and $\kappa(\bx)$. We observe that the results are qualitatively similar to those in Fig. \ref{fig:Decomps_cstMu_2DRTI}, and that differences between the three decompositions are somewhat enhanced. 
We also repeated the `R4vc' case at a lower density ratio $\rho_h/\rho_l=100$ with similar results (not shown here).

\clearpage
\newpage

\subsection{3D Rayleigh-Taylor Instability\lb{sec:3DRTI}}
Equations (\ref{continuity})-(\ref{total-energy}) with $g=1$  are used to conduct a simulation of a Rayleigh-Taylor instability (RTI) in 3D using our code DiNuSUR. We use no-slip BC at the top and bottom walls and periodic BC in the horizontal directions. 
The domain is $L_x\times L_y\times L_z=1.6\times 1.6\times 3.2$. We use a  $N_x\times N_y\times N_z= 512\times 512\times 1{,}024$ grid, a pseudospectral solver in the horizontal direction and a 6th-order compact finite difference scheme in the z-direction. The initial conditions of the simulations are those of a dense fluid with $\rho_h=1.0$, filling the top-half of the domain in the z-direction, over a less dense fluid with $\rho_l=1/9$ in the bottom half. As in the 2D RT cases, the initial pressure satisfies hydrostatic equilibrium $dP/dz=-\rho g$. Small amplitude velocity perturbations result in RTI which evolves into the fully turbulent regime. Dynamic viscosity, $\mu=1.35\times 10^{-4}$, is taken as a constant, although we have shown above that our results pertaining to the inviscid criterion also hold if $\mu$ has significant spatial variations. 

\begin{figure}
{\includegraphics[height=3in]{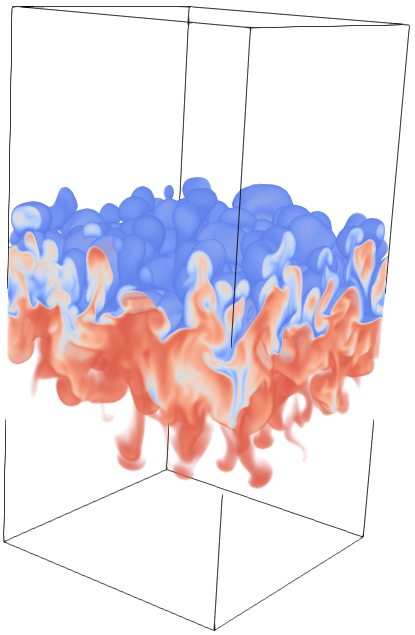}} 
\caption{\footnotesize{Visualization of density in the Rayleigh-Taylor Instability flow at the instant of time we use in our analysis.} 
\label{fig:RTviz37000}}\end{figure}

The estimated Reynolds number of this RT flow is $Re=\frac{L_x}{\nu}\sqrt{\frac{AgL_x}{1+A}}=9943$, the integral length scale is the largest scale that gravity acts on, which is the domain size $L_x=1.6$. The mesh Grashof number is $Gr=2Ag\langle \rho\rangle^2 \Delta^3/\mu^2=0.822$. The Kolmogorov length scale $\eta$ is $\eta=\frac{\mu^{3/4}}{\epsilon^{1/4}\langle \rho\rangle^{3/4}}=7.681\times 10^{-3}$, where $\epsilon=4.157\times 10^{-3}$ is the specific energy dissipation rate. In contrast, the grid size is $\Delta x=0.0031$, which gives $\eta/\Delta x = 2.478$. A visualization of density at the time we analyze the flow is shown in Fig.~\ref{fig:RTviz37000}.

\begin{figure}
{\includegraphics[height=3in]{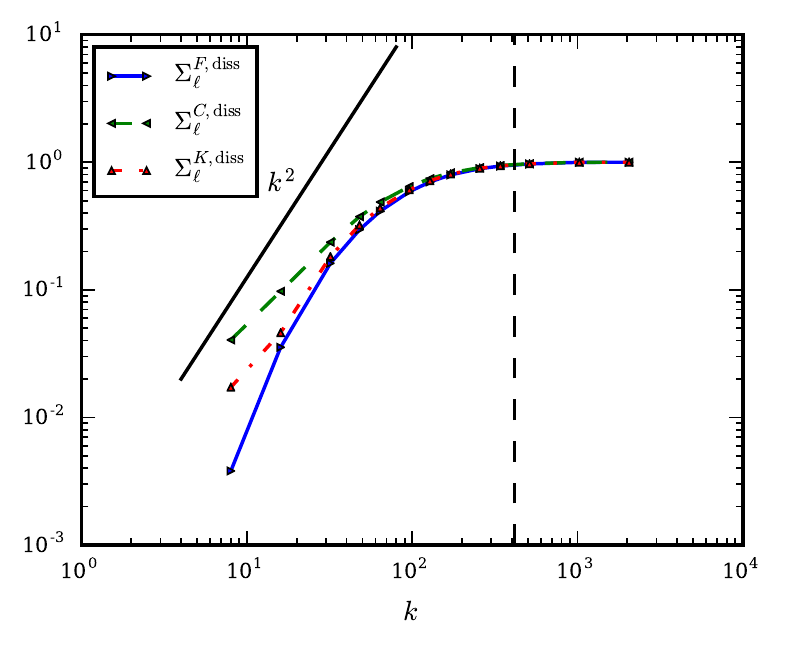}} 
\caption{\footnotesize{$L^1$-norm scaling of the viscous dissipation, $\|\Sigma^{\mbox{\scriptsize{diss}}}_\ell\|_1$ using the original 3D RT field in Fig. \ref{fig:RTviz37000}. Plot is normalized by the unfiltered dissipation. In the horizontal axis $k=L_z/\ell$,  and the vertical line marks the viscous cut-off wavenumber $k_d=L_z/\eta$. The Favre dissipation decays the fastest, at least as $\ell^{-2}$ at large scales.}
\label{fig:3Ddissip_loglog}}\end{figure}

With this data, we analyze the viscous contribution from each of the three decompositions. Fig.~\ref{fig:3Ddissip_loglog} plots the mean magnitude of dissipation corresponding to three scale decompositions, and shows that the Favre definition yields the fastest decay of dissipation at large scales. 

\begin{figure*}[bhp]
\begin{minipage}[b]{1.0\textwidth}  
\subfigure[\footnotesize{Density spectra}]
{\includegraphics[width=3in]{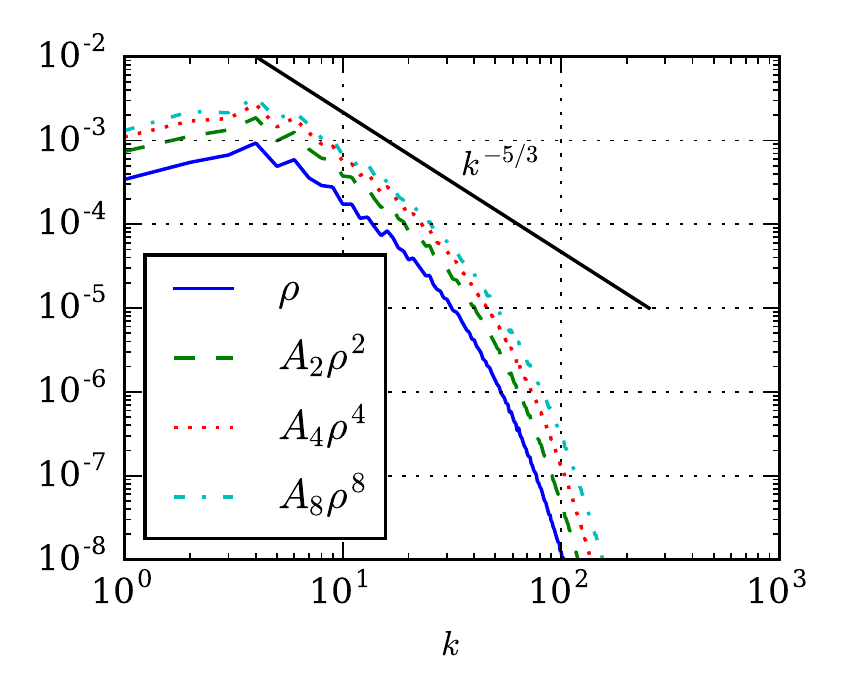}} 
\phantom{a}
\subfigure[\footnotesize{Density PDF}]
{\includegraphics[width=3in]{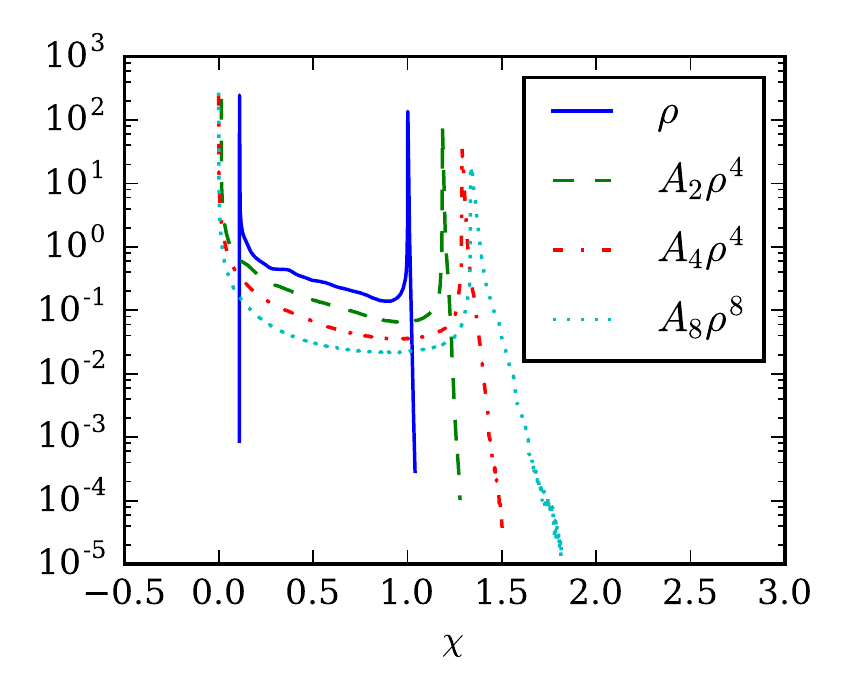}}
\caption{\footnotesize{The spectra and pdf of density fields from the snapshot shown in Fig. \ref{fig:RTviz37000}. Spectra are calculated by a 2D Fourier transform in the horizontal directions and then averaging along the z-direction, horizontal axis $k=L_z/\ell$. Density spectra in Cases D1-D8 have similar scaling but with different fluctuation intensity.} \label{fig:DensityPDF37000}}
\end{minipage}
\end{figure*}

While our flow has significant density contrast, with an initial ratio of $\rho_h/\rho_l = 9$, achieving higher ratios in a well-resolved turbulence simulation in 3D is computationally challenging (e.g. \cite{livescu2008variable}). As we mentioned, many flows of interest have very large density ratios. Since the inviscid criterion is a kinematic result, independent of the dynamics as we have discussed in section \ref{sec:InviscidCriterion} above, and in order to highlight differences in the kinematic (or functional) behavior of the viscous terms, eqs. (\ref{eq:FavreVD})-(\ref{eq:KritsukVD}), arising from the three decompositions under higher density contrast, we synthetically increase the density contrast in the flow we are analyzing by taking powers of the density $\rho^m(\bx)$ with $m =2,4,8$, as a post-processing step. 
$\rho^m$ is then normalized such that the total mass in the domain is the same as in the original flow, $\langle\rho\rangle = A_{m}\langle\rho^m\rangle$. We then use the three synthetic density fields, $\chi=A_m\rho^m$, to calculate the terms in eqs. (\ref{eq:FavreVD})-(\ref{eq:KritsukVD}). Table \ref{tbl:DensityCases} summarizes the four cases we consider and Fig. \ref{fig:DensityPDF37000} shows the spectra and probability density function (pdf) of the four density fields. The spectra of the three synthetic density fields are physically reasonable in the sense that they are very similar to the spectrum of the original data, although spatial correlations of $\chi$ with dynamically relevant fields (e.g. pressure or vorticity) need not be.
This justifies using these synthetic density fields to test for the inviscid criterion at the kinematic (or functional) level.

\begin{table} \centering
\setlength{\tabcolsep}{4pt}
\caption{We consider four cases: the RTI flow shown in Fig. \ref{fig:RTviz37000} with the original density field, $\rho$, along with three cases, D2-D8, using synthetic density fields to amplify density gradients. Here, $\chi'_{rms} = \langle(\chi-\langle\chi\rangle)^2\rangle^{1/2}$.}
\begin{tabular}{|c| c |c|c|c|}
\hline  &Density $\chi$ & $\langle\chi\rangle$ & $\chi'_{rms}$ & $\frac{\chi'_{rms}}{\langle \chi\rangle}$   \\ \hline
D1  & $\rho$ & 0.554 & 0.405 & 0.731 \\ \hline
D2 & $A_{2}\rho^2$ & 0.554 & 0.546 & 0.986 \\ \hline
D4  & $A_4\rho^4$ & 0.554 & 0.619 & 1.118 \\ \hline
D8   & $A_8\,\rho^8$ & 0.554 & 0.654 & 1.182 \\ \hline
\end{tabular}
\label{tbl:DensityCases}
\end{table}

\begin{figure*}[bhp]
\begin{minipage}[b]{1.0\textwidth}  
\subfigure[\footnotesize{$\|\Sigma_\ell^{F,\text{diss}}\|_1$}]
{\includegraphics[height=1.7in]{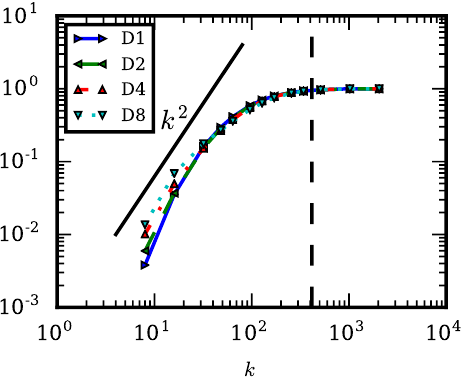}} 
\phantom{}
\subfigure[\footnotesize{$\|\Sigma_\ell^{C,\text{diss}}\|_1$}]
{\includegraphics[height=1.7in]{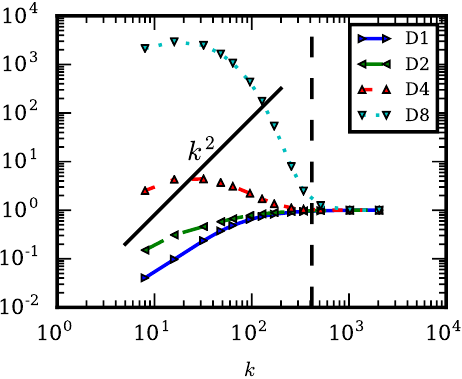}}
\phantom{}
\subfigure[\footnotesize{$\|\Sigma_\ell^{K,\text{diss}}\|_1$}]
{\includegraphics[height=1.7in]{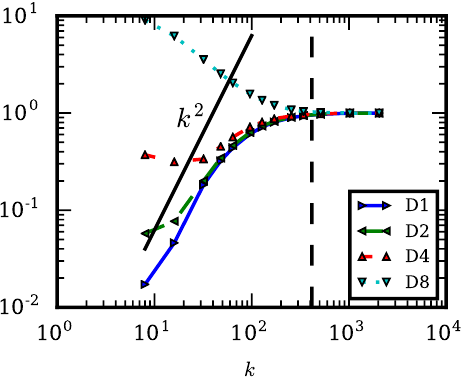}}\\
\caption{\footnotesize{Sensitivity to increasing density variations: log-log plot of the $L^1$-norm of dissipation for each decompostion. Horizontal axis  are $k\equiv L_z/\ell$, and the vertical line in each figure marks the viscous cut-off wavenumber $k_d=L_z/\eta$. This shows that the Favre dissipation term decays at least as fast as $\ell^{-2}$ regardless of the intensity of density variability, unlike the other two decompositions. Note that plots of case D1 are the same as in Fig. \ref{fig:3Ddissip_loglog}.
\label{fig:37000const_loglog}}}
\end{minipage}
\end{figure*}

Fig. \ref{fig:37000const_loglog} shows the sensitivity of $\Sigma_\ell$ from the three decompositions to increasing density variations. We observe that the the Favre decomposition satisfies the inviscid criterion and $\|\Sigma_\ell^{F,\text{diss}}\|_1$ decays at least as fast as $\ell^{-2}$ for all density fields considered, regardless of the intensity of density variations, in agreement with the results in \cite{Aluie13}. On the other hand, we can clearly see in Fig.~\ref{fig:37000const_loglog} that viscous terms in the non-Favre decompositions exhibit a strong sensitivity to density variations. In the presence of strong density variation, $\|\Sigma_\ell^{C,\text{diss}}\|_1$ and $\|\Sigma_\ell^{K,\text{diss}}\|_1$ do not decay at large scales, in violation of the inviscid criterion. We believe that the absence of such a stark sensitivity to density variations in cases R2-R4 of the previous subsection \ref{sec:2DRTI} was probably due to the low level of turbulence in those 2D flows, as we have remarked earlier.

\clearpage
\newpage

\section{Summary}
We analyzed the viscous contribution at different `length-scales' of several flows in one-, two-, and three-dimensions, and showed that not all scale-decompositions are equivalent. In the presence of significant density variations, a Favre (or Hesselberg) decomposition satisfies the inviscid criterion by guaranteeing that viscous effects are negligible at large `length-scales' regardless of the intensity of density fluctuations as was shown mathematically in \cite{Aluie13} and demonstrated numerically here. 

We also showed how two non-Favre decompositions commonly used in the literature yielded viscous contributions several orders of magnitude greater than that of Favre at `large-scales.' Our results also suggest that these viscous effects may not decay at large length-scales in some of the flows we considered, in violation of the inviscid criterion. Therefore, these non-Favre decompositions are not appropriate to analyze inertial-range dynamics in the presence of significant density variations. This has important bearings on attempts to study the energy transfer in variable density turbulence using ``triadic interactions'' or using $\sqrt{\rho}\bu$ as the elemental variable (e.g. \cite{CookZhou02,Toweryetal16,PraturiGirimaji17,Greteetal17}). While triadic interactions are appropriate in incompressible turbulence, where the energy transfer nonlinearity is cubic, they may not be valid for studying energy transfer in variable density turbulence since the scale-decomposition associated with such a triadic analysis may not satisfy the inviscid criterion. We remark that the observation of putative power-law scalings of a quantity, such as $|\widehat{\sqrt{\rho}\bu}|^2(k)$, is not sufficient to infer that the quantity is undergoing an inertial-range cascade. The results of this paper also have practical modeling implication in showing that viscous terms in Large Eddy Simulations do not need to be modeled and can be neglected if the resolved scales are large enough.

\clearpage
\newpage

\begin{acknowledgments}
We thank two anonymous referees for valuable comments that helped improve the manuscript. This work was supported by the DOE Office of Fusion Energy Sciences grant DE-SC0014318 and the DOE National Nuclear Security Administration under Award DE-NA0001944. HA was also supported by NSF grant OCE-1259794 and by the LANL LDRD program through project number 20150568ER. An award of computer time was provided by the INCITE program, using resources of the Argonne Leadership Computing Facility, which is a DOE Office of Science User Facility supported under Contract DE-AC02-06CH11357.
This research also used resources of the National Energy Research Scientific Computing Center, a DOE Office of Science User Facility supported by the Office of Science of the U.S. Department of Energy under Contract No. DE-AC02-05CH11231. 

\end{acknowledgments}

\section*{Appendix}
\begin{figure*}
\begin{minipage}[b]{1.0\textwidth}  
\subfigure[\footnotesize{Dissipation average}]
{\includegraphics[height=2.6in]{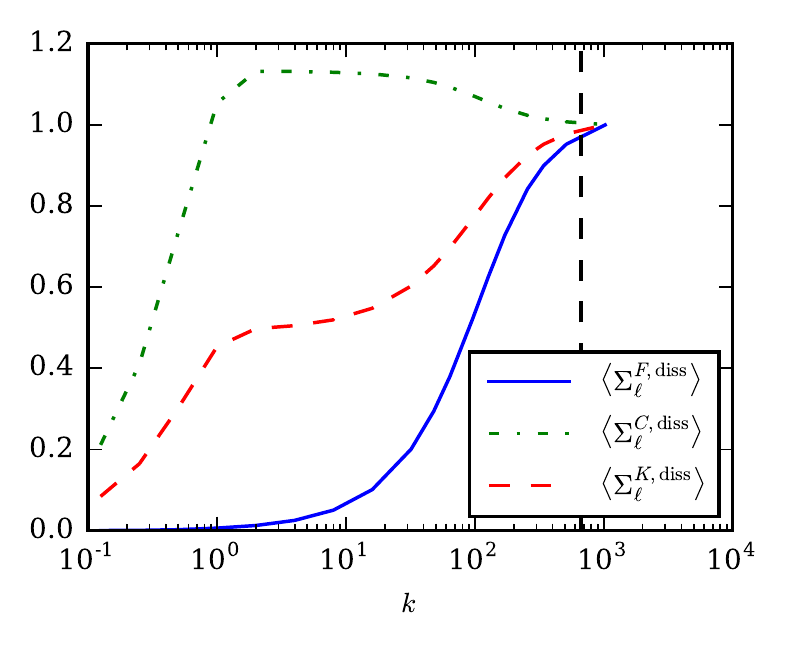}} 
\phantom{a}
\subfigure[\footnotesize{Dissipation $L^\infty$ log-log}]
{\includegraphics[height=2.6in]{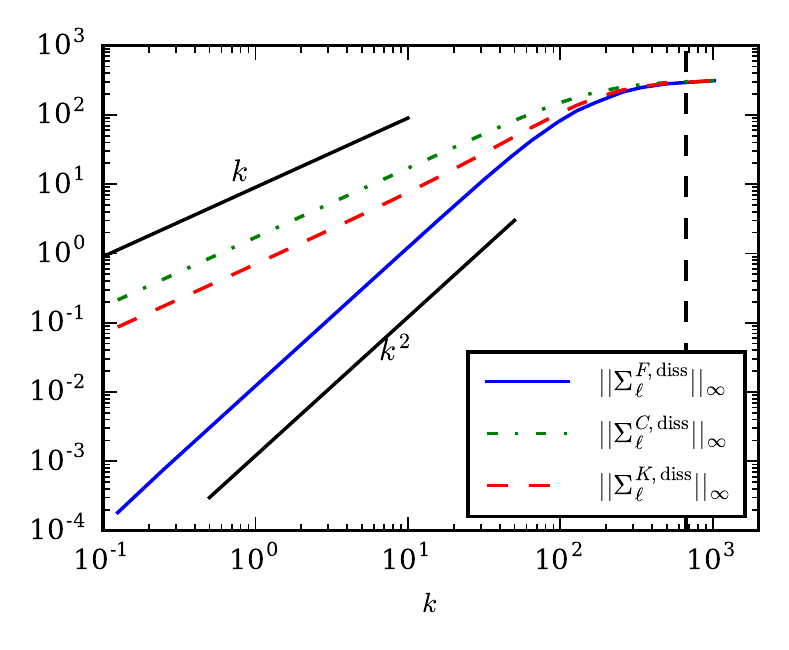}}
\caption{\footnotesize{Ma 3 constant spatial viscosity results. Left: average dissipation, right: $L^\infty$ of dissipation. Data is normalized by the unfiltered dissipation, and the vertical line in each figure marks the viscous cut-off wavenumber $k_d=L/\ell_\mu$.} \label{fig:appendix_shock}}
\end{minipage}
\end{figure*}

In Figure \ref{fig:appendix_shock}, we show results from a 1D normal shock case identical to the M3 case, but with a spatially constant viscosity. The results are very similar to those in Fig. \ref{fig:Ma1.2_dissip_stat} above. This is consistent with our previous assertions that our conclusions are independent of whether or not $\mu$ is spatially varying.

\begin{figure}
{\includegraphics[height=3in]{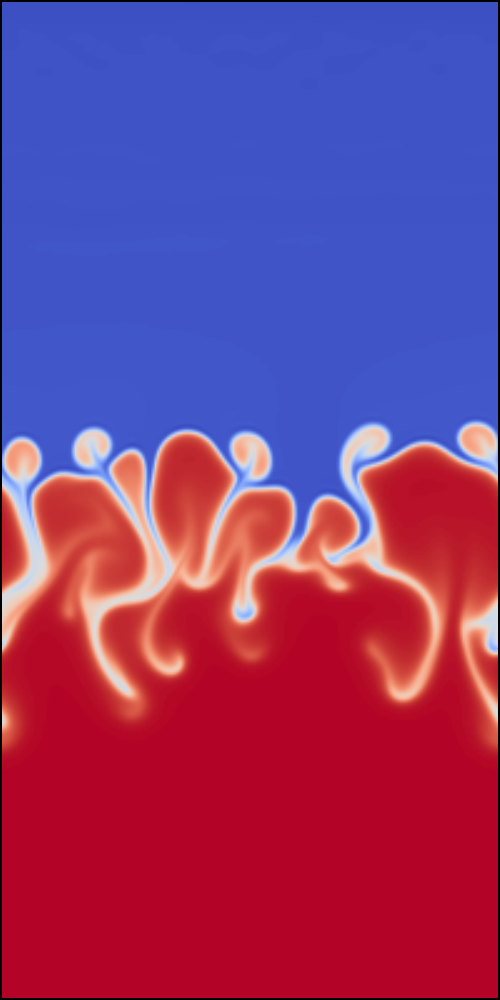}} 
\caption{\footnotesize{Visualization of density in the R3 case of 2D RT flow at later time.} 
\label{fig:appendix_2Dviz}}\end{figure}

\begin{figure*} 
\begin{minipage}[b]{1.0\textwidth}  
\subfigure[\footnotesize{R3: $\langle\Sigma_\ell\rangle$}]
{\includegraphics[height=2.6in]{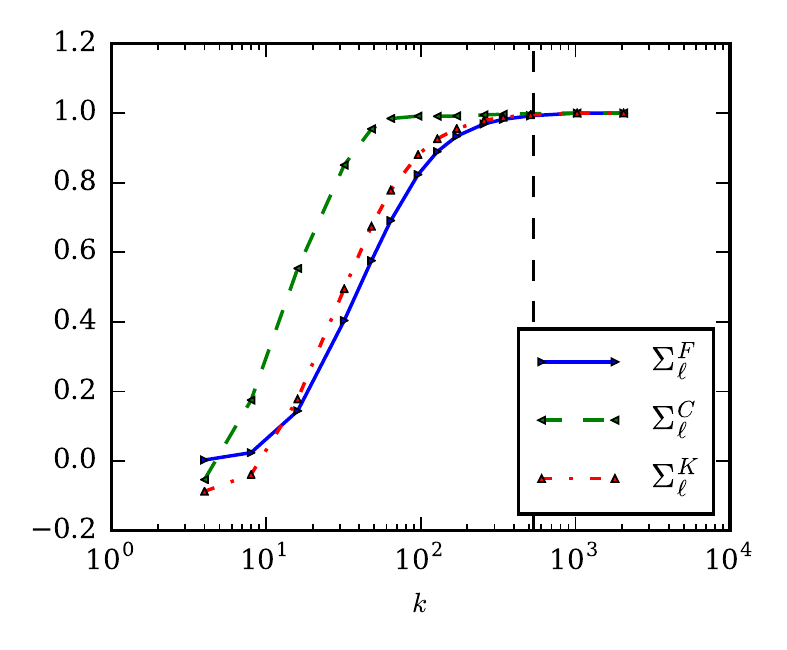}} 
\phantom{a}
\subfigure[\footnotesize{R3: $\|\Sigma^{\mbox{\scriptsize{diss}}}_\ell\|_1$}]
{\includegraphics[height=2.6in]{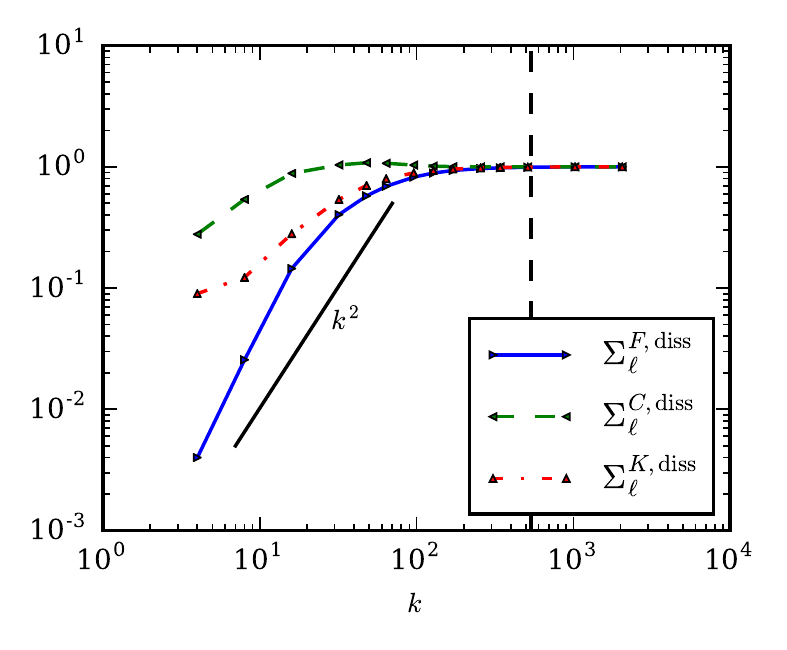}}
\caption{\footnotesize{The viscous contribution and viscous dissipation of R3 at a later time than that used in Fig.~\ref{fig:Decomps_cstMu_2DRTI} above. Left: average viscous contribution, right: $L^1$ norm of dissipation. Data is normalized by the unfiltered dissipation, and the vertical line in each figure marks the viscous cut-off wavenumber $k_d=L_z/\eta$.} \label{fig:appendix_2DRT}}
\end{minipage}
\end{figure*}

In Figure \ref{fig:appendix_2Dviz} and \ref{fig:appendix_2DRT}, we show results from the R3 case of 2D RT flow, but at a later time at which the mixing height (average height between bubble and spike) is $\approx 1.5$ times that in Fig. \ref{fig:Viz_2DRTI}, as is visualized in Fig.\ref{fig:appendix_2Dviz}. The results are very similar to those in Fig. \ref{fig:Decomps_cstMu_2DRTI} above, showing that the particular snapshots we chose to analyze above in the RT flows are not special and that our results hold in general.

\bibliographystyle{unsrt}
\bibliography{scale-citation}

\end{document}